\documentclass[%
 reprint,
superscriptaddress,
 amsmath,amssymb,
 aps,
prb,
]{revtex4-1}

  \usepackage{float} 
  \usepackage[utf8]{inputenc}                  
  \usepackage[english]{babel}                  
  \usepackage[T1]{fontenc}                     
  \usepackage{lmodern}
  \usepackage{dcolumn}                         
  \usepackage{graphicx}                        
  \usepackage[svgnames]{xcolor}                
  \usepackage{amsmath}
  \usepackage{amsfonts}
  \usepackage{amssymb}
  \usepackage{mathtools}
  \usepackage{bm}
  \usepackage{nicefrac}
  \usepackage{dsfont}
  \usepackage{mathrsfs}
  \usepackage{multirow}                        
  \usepackage{tabularx}                        
  \usepackage{longtable}                       
  \usepackage[breaklinks=true]{hyperref}
  \usepackage{cleveref}
  \usepackage[]{siunitx} 

\newcommand{\abs}[1]{\ensuremath{ \left| #1 \right| }}                    
\newcommand{\mean}[1]{\ensuremath{ \left\langle#1\right\rangle }}         

\newcommand{\dd}[0]{\mathrm d}                
\newcommand{\del}[0]{\partial}                

\renewcommand{\rho}[0]{\varrho}
\renewcommand{\theta}[0]{\vartheta}
\renewcommand{\phi}[0]{\varphi}

\renewcommand{\vec}[1]{\bm{#1}}        

\newcommand{\ba}[1]{\begin{align} #1 \end{align}}

\newcommand{\muS}{\mu_s}

\newcommand{\kBT}{k_\mathrm{B}T}
\newcommand{\Neel}{N\'{e}el}

\begin{document}


\title{Magnonic Proximity Effect in Insulating Ferro- and Antiferromagnetic Trilayers}

\author{Verena Brehm}
  \email{verena.brehm@uni-konstanz.de}
  \affiliation{%
  	Fachbereich Physik, Universit\"at Konstanz, 78457 Konstanz, Germany
  }%
\author{Martin Evers}%
  \affiliation{%
    Fachbereich Physik, Universit\"at Konstanz, 78457 Konstanz, Germany
  }%
\author{Ulrike Ritzmann}%
  \affiliation{%
    Fachbereich Physik, Freie Universit\"at Berlin, 14195 Berlin, Germany
  }%
\author{Ulrich Nowak}
\affiliation{%
  Fachbereich Physik, Universit\"at Konstanz, 78457 Konstanz, Germany
}%

\date{\today}

\begin{abstract}
The design of spin-transport based devices such as magnon transistors or spin valves will require multilayer systems composed of different magnetic materials with different physical properties.
Such layered structures can show various interface effects, one class of which being proximity effects, where a certain physical phenomenon that occurs in the one layers leaks into another one. 
In this work a magnetic proximity effect is studied in trilayers of different ferro- and antiferromagnetic materials within an atomistic spin model. We find the magnetic order in the central layer -- with lower critical temperature -- enhanced, even for the case of an antiferromagnet surrounded by ferromagnets. We further characterize this proximity effect via  the magnon spectra which are specifically altered, especially for the case of the antiferromagnet in the central layer.

\end{abstract}

\maketitle


\section{Introduction}
Spintronics is based on the increasing efforts to replace or supplement electronic devices by devices that exploit spin-transport phenomena.
Especially, in magnonic devices one would try to avoid charge currents, utilizing magnons -- the elementary excitations of a magnets ground state -- for the spin transport
\cite{Chumak14_MagnonTransistor,Klinger14_SpinWaveLogicDevices,Kruglyak10_MagnonicsReview,Chumak15_MagnonSpintronics}.
The great potential of this idea has been demonstrated, for instance, by the magnon transistor \cite{Chumak14_MagnonTransistor}, which forms a building block for magnon-based logic \cite{Klinger14_SpinWaveLogicDevices}.
A further development in this context is the use of antiferromagnets \cite{Lebrun18_LongDistTransportHematite,Khymyn16_TransfromSpincurrentByAFMinsulator} which, for instance, allow to build a spin-valve structure \cite{Cramer18_SpinValve} -- a multilayer system designed to pass spin waves through the central, antiferromagnetic layer only, when the two outer, ferromagnetic layers are magnetized in the same direction.
For an antiparallel magnetization, the magnons are blocked.

A variety of spin-transport experiments in antiferromagnets is to monochromatically pump spin waves from a ferromagnet via ferromagnetic resonance into an antiferromagnet and detect the signal via the inverse spin-Hall effect in an attached heavy metal layer \cite{Wang14_AFMSpinTransportYIGintoNiO,Wang15_FMRexcitationOfAFMinsulators,Qiu2016_SpinCurrentProbeForAFMPhaseTransitions}.
Alternatively, one can excite thermal spin waves via the spin-Seebeck effect \cite{Lin16_SSEthroughAFMinsulator,Prakash16_SSEthroughNiO,Cramer18_SEEAcrossAFM}. 

In either way, the setup in total is necessarily a trilayer system, where two layers are magnetically ordered and the two materials may have different ordering temperatures. 
This raises the question for the temperature dependence of the spin transport, especially when the two critical  temperatures are quite different and proximity effects at the interface may play a role.
First experiments in such systems exist, including temperature ranges well above the critical temperature of one of the constituents. Surprisingly, even then there seems to be a spin current above the \Neel{} temperature of the antiferromagnet, as demonstrated e.g.\ in \cite{Cramer18_SEEAcrossAFM,Goennenwein_2018_MRatNeelTemp}.

For a deeper understanding of the temperature dependent magnetic behavior of these multilayer systems, it is necessary to study the impact of one layer onto the magnetic behavior of the other, a class of effects that is called magnetic proximity effect  \cite{Manna14_ReviewExchangeBiasAndMagneticProximityEffect}. 
Magnetic proximity effects have been investigated in bilayers composed of an itinerant ferromagnet coupled to a paramagnet, where magnetic moments are induced in the paramagnet \cite{Zuckermann73_MagneticProximityEffect,Cox79_MagnProximityEffectItinerantFM,Mata82_ModelMagnProximityEffectItinerantFMfilms}, but it is rather ubiquitous for all kinds of heterostructures and also core-shell nanoparticles \cite{Carey93_InterlayerCouplingCoONiO,Borchers93_InterlayerCouplingCoONiO,Lenz07_MagnProximityAFM_FM_Bilayer,Maccherozzi08_MagnProximity_Fe_SemiconductorInterface,Golosovsky09_MagnProximityEffect_FM_AFMCoreShell}.
Typical signatures of proximity effects are a magnetization in a paramagnetic constituent, an enhanced ordering temperature in the material with the lower ordering temperature, an increased coercivity, and also the occurrence of an exchange bias effect.\ \cite{Manna14_ReviewExchangeBiasAndMagneticProximityEffect}

Theoretically, proximity effects in bilayers of ferro- and antiferromagnets have been investigated using mean-field techniques \cite{Jensen05_TheroyMagnProximityEffect_FM_AFM} , Monte Carlo simulations \cite{nowakPRB02}, and multi-scale techniques \cite{szunyoghPRB11}. However, these studies neglect the influence of magnons that might pass the interface of a magnetic bilayer as these can only be treated via spin dynamics calculations. It is, hence, the purpose of this study to investigate the magnetic proximity effect including magnon spectroscopy, with that adding to a more  complete understanding of the temperature dependent magnetic behavior of bi- and trilayers close to the interface. 

The outline of this work is as follows: in \cref{sec:ASM} we describe our model and the two setups which we treat in the following -- a magnetic trilayer system build up of three ferromagnets, where the central layer has a lower Curie temperature, and a corresponding ferromagnet-antiferromagnet-ferromagnet system.
We investigate the temperature-dependence of the spatially resolved order parameters and susceptibility in \cref{subsec:FM-FM-FM,subsec:FM-lAFM-FM}, and the magnon spectra in \cref{subsec:MagnonSpectra}.
We show that each property can probe this proximity effect, especially in the vicinity of the critical temperature of the central layer and discover a magnonic contribution to the proximity effect that rests on the different spectra and polarizations of magnons in the different layers.


\section{Model, methodology and geometry}  \label{sec:ASM} 

We conduct our work within an atomistic spin model, where every magnetic atom at position $\vec{r}_l$, $l = 1,...,N$, is described by a classical magnetic moment $\vec{\mu}_l = \muS\vec{S}_l$ of magnitude $|\vec{\mu}_l | = \muS$.
Assuming a model of Heisenberg type, the Hamiltonian of the system reads
\begin{align}
  \operatorname{H} = -\frac{1}{2} \sum_{ \mathclap{ \substack{j,k = 1 \\ k\in \mathrm{NN}(j)} } }^{N} J_{jk} \vec{S}_j\cdot \vec{S}_k - d_z\sum_{j=1}^{N} S_{j,z}^2  
\end{align}
with the Heisenberg exchange interaction $J_{jk}$, restricted to nearest neighbors (NN), and a uniaxial anisotropy, parameterized by the anisotropy constant $d_z > 0$.
The equation of motion is the Landau-Lifshitz-Gilbert equation \cite{Landau35_LL_equation} with Gilbert damping $\alpha$ \cite{Gilbert55_Gilbert_damp,Gilbert04_Gilbert_damp_IEEE},
\begin{align}
  \partial_t \vec{S}_l & = -\frac{\gamma}{\muS(1+\alpha^2)}\left[ \vec{S}_l\times \vec{H}_l + \alpha\vec{S}_l \times (\vec{S}_l \times \vec{H}_l) \right]
 \end{align}
  with gyromagnetic ratio $\gamma > 0$ and the effective field
\begin{align}  
  \vec{H}_l            & = -\frac{\partial \operatorname{H}}{\partial \vec{S}_l} + \vec{\xi}_l .
\end{align}
The coupling to the heat bath at temperature $\kBT$ \cite{Brown63_ThermalFluctuactionsMagnParticles} leads to thermal fluctuations in form of a Gaussian white noise $\vec{\xi}$ satisfying $\mean{ \vec{\xi}_l }=0$ and 
\begin{align}
  \mean{ \xi_{l,\beta}(t) \xi_{k,\zeta}(t') } & = \frac{2 \muS \alpha \kBT} {\gamma}\delta_{lk}\delta_{\beta\zeta}\delta(t - t')                                 
\end{align}
with  $\beta,\zeta  \in \{ x,y,z \} $. These stochastic differential equations are solved numerically using the stochastic Heun algorithm. \cite{Nowak07_SpinModels}
The simulations are implemented in a highly efficient code developed in \textsc{C/C++} and \textsc{CUDA}, running on GPUs.  A high degree of optimization is necessary because of the rather large system size (about $N \approx 10^5$ spins) in combination with very long equilibration times close to the critical temperature.

\begin{figure}
  \centering
  \includegraphics[width=1.0\linewidth]{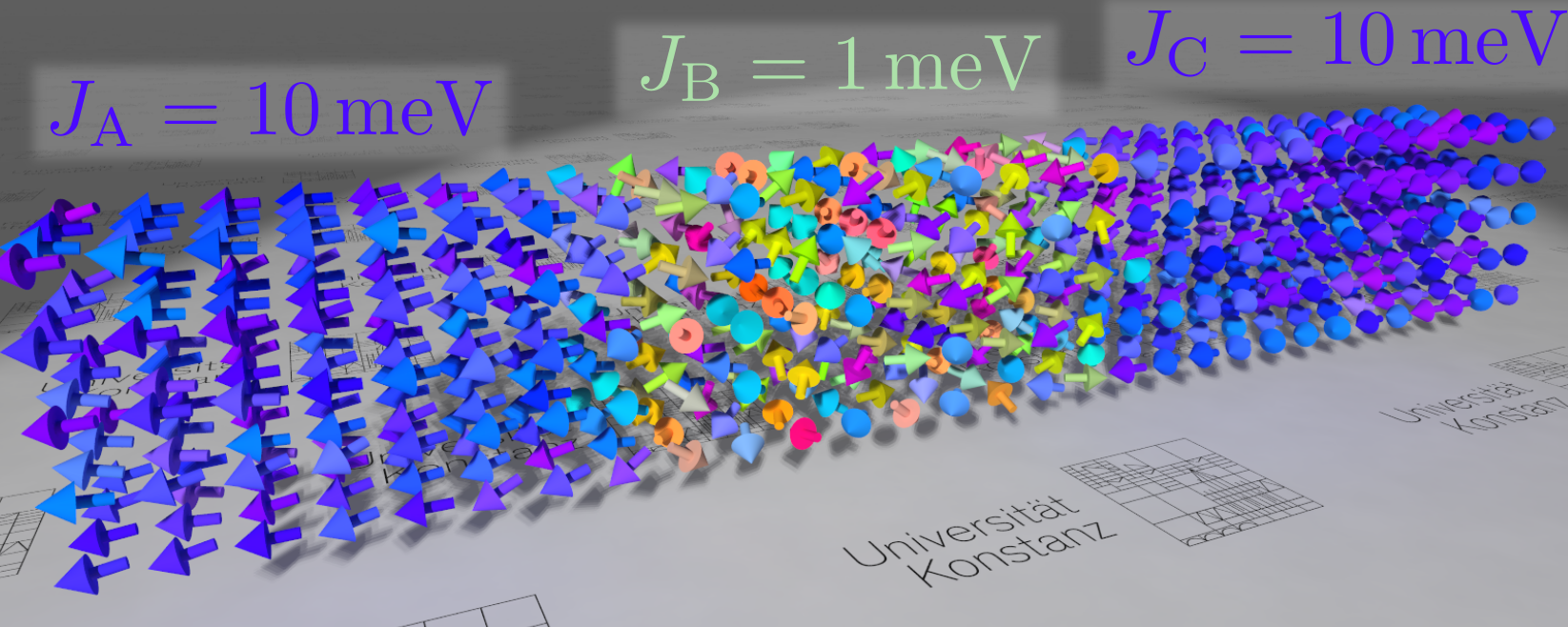}
  \caption{Geometry of the investigated trilayer: in the central layer a lower exchange constant $J_\mathrm{B}$ is used, leading here to  a much lower critical temperature $T_\mathrm{c}$ than in the outer layers.}
  \label{fig:ExchangeTrilayer}
\end{figure}

The system of interest is a trilayer stacked along the $z$ direction -- the three layers denoted A, B and C -- composed of spins arranged on a simple cubic lattice with lattice constant $a$, see \cref{fig:ExchangeTrilayer}.
The Heisenberg coupling constant varies along the system by a factor of $10$: within each layer it takes isotropic values $J_{jk} = J_\mathrm{A},\pm J_\mathrm{B},J_\mathrm{C}$, where $10 J_\mathrm{B} = J_\mathrm{A} = J_\mathrm{C}$. This choice results in very different critical temperatures.
At the interfaces we choose a coupling of intermediate strength $J_{jk} = \nicefrac{J_\mathrm{A}}{2}$ for lattice sites $j,k$ at the interfaces of layers A and B as well as layers B and C.
There are two different setups: a purely ferromagnet trilayer (termed FM-FM-FM) with $J_{\mathrm{B}} > 0$, and a layered antiferromagnet sandwiched between two ferromagnets (denoted FM-lAFM-FM).
In the latter case, the exchange is ferromagnetic, $J_{lk} = J_\mathrm{B} > 0$, in the $x$-$y$ plane and antiferromagnetic along the $z$ direction, $J_{lk} = -J_\mathrm{B} < 0$.
The use of the layered antiferromagnet ensures the interfaces to be ideal in either case (parallel alignment of the spins in the ground state), corresponding to completely uncompensated interfaces.\\
As a test case, we choose the following values for our model parameters: $J_\mathrm{A} = J_\mathrm{C} = \SI{10}{\milli\electronvolt}$, $J_\mathrm{B} = \SI{1}{\milli\electronvolt}$ and $d_z = \SI{0.1}{\milli\electronvolt}$. 
Furthermore, it is $\gamma = \gamma_0$, the free electron's gyromagnetic ratio, and $\mu_s = \mu_\mathrm{B}$, Bohr's magneton.

\section{Results}
We study the magnetic trilayers outlined above with respect to the magnetic proximity effect in terms of three aspects: their temperature-dependent order parameter profiles, their temperature-dependent susceptibility profiles, and their magnon spectra. 

\subsection{Magnetization of a Ferromagnetic Trilayer} \label{subsec:FM-FM-FM} 
In a first step, we consider the equilibrium order parameter profile along the $z$ direction.
For a ferromagnet this is the magnetization, which we resolve monolayer-wise along $z$ direction,
\ba{
  \mean{ m_z }(z) = \frac{1}{N_{xy}}\sum_{r_{j,z} = z} \langle S_{j,z} \rangle,
}
with $N_{xy}$ being the number of spin per monolayer and $\mean{\ldots}$ denoting the thermal average, which we calculate in our simulations as time average.

\begin{figure}
  \centering
  \includegraphics[width=1.0\linewidth]{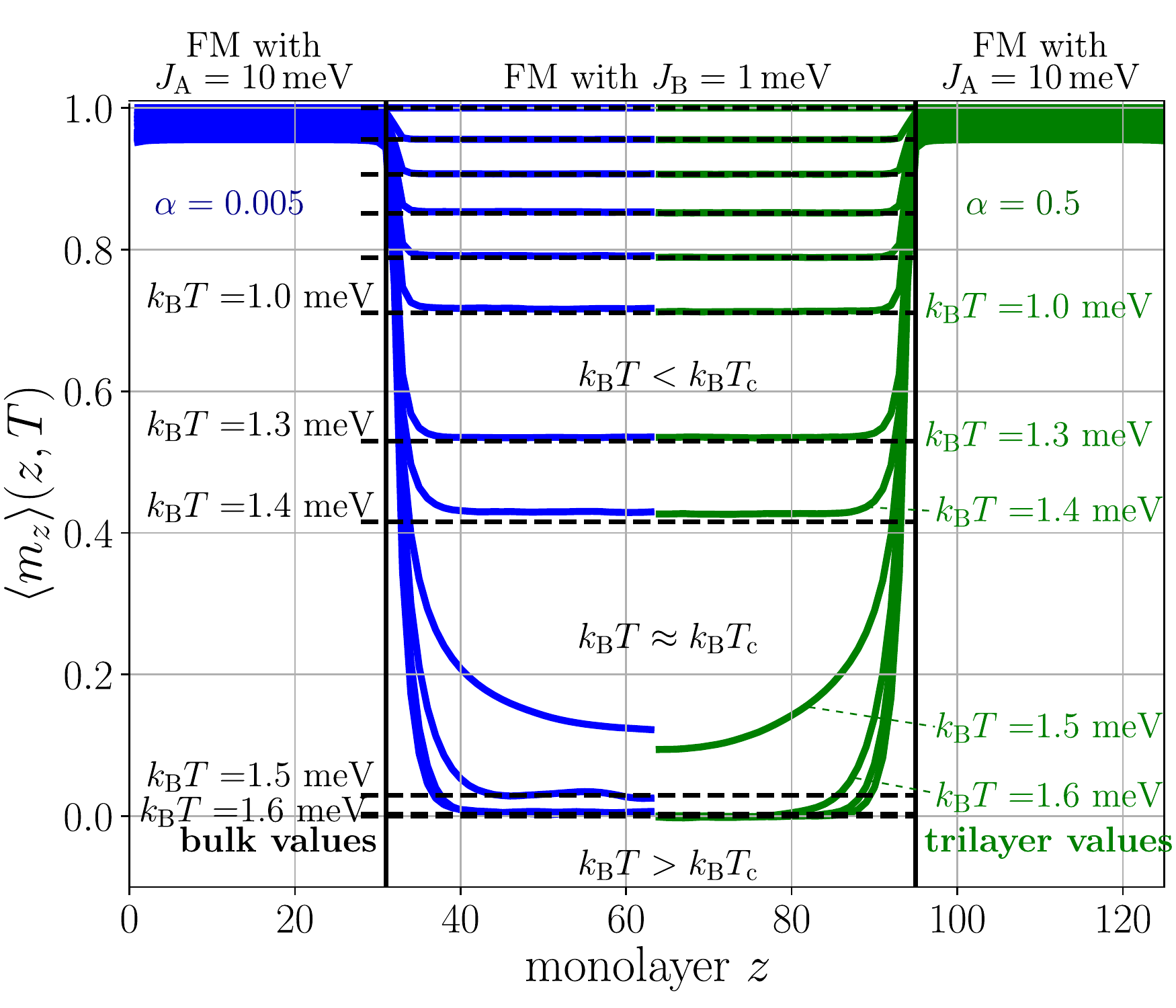}
  \caption{%
    Magnetization profiles along $z$ axis in a FM-FM-FM trilayer with coupling ratio $\nicefrac{J_\mathrm{B}}{J_\mathrm{A}} = \num{0.1}$ for different temperatures. 
    To illustrate the influence of the damping constants we present in the left part our results for $\alpha = \num{0.005}$ (blue) and in the right part only data for $\alpha = \num{0.5}$ (green).
    The bulk value of the equilibrium magnetization with coupling constant $J_\mathrm{B}$ is shown as black dashed lines for comparison.
    The corresponding critical temperature is $\kBT_\mathrm{c} \approx \SI{1.5}{meV}$. 
  }
  \label{fig:FM-mz}
\end{figure}

\Cref{fig:FM-mz} depicts this magnetization for the FM-FM-FM system. Vertical lines indicate the interfaces at $z = 32a$ and $z = 64a$, separating the central layer B with low exchange constant $J_\mathrm{B}$ from the outer layers A and C with a coupling constant that is ten times higher. We tested two very different values of the damping constants $\alpha = 0.005$ and $\alpha = 0.5$, corresponding to the blue and green lines in the figure respectively.
There is only a small difference visible close to the transition temperature $\kBT_\mathrm{c} \approx 1.5J_\mathrm{B}$, where the curve for the smaller damping is not fully converged to its equilibrium profile. We conclude that -- apart from this small deviation -- our results are equilibrium properties, that do not depend on $\alpha$.

The outer layers show a rather stable magnetization with respect to an increasing temperature due to the higher exchange constant, while the central layer undergoes a phase transition where the magnetization drops to zero.
However, there is a significant difference to a bulk material with exchange constant $J_\mathrm{B}$ (indicated as black dashed lines): while for low temperatures magnetization values of the bulk and the central layer of the trilayer are in good agreement, the magnetization of the central layer remains significantly higher in the vicinity of the critical temperature. Especially, there is a residual magnetization in the central layer directly at the critical temperature, a first  signature of a magnetic proximity effect.

\begin{figure}
  \centering
 \includegraphics[width=0.9\linewidth]{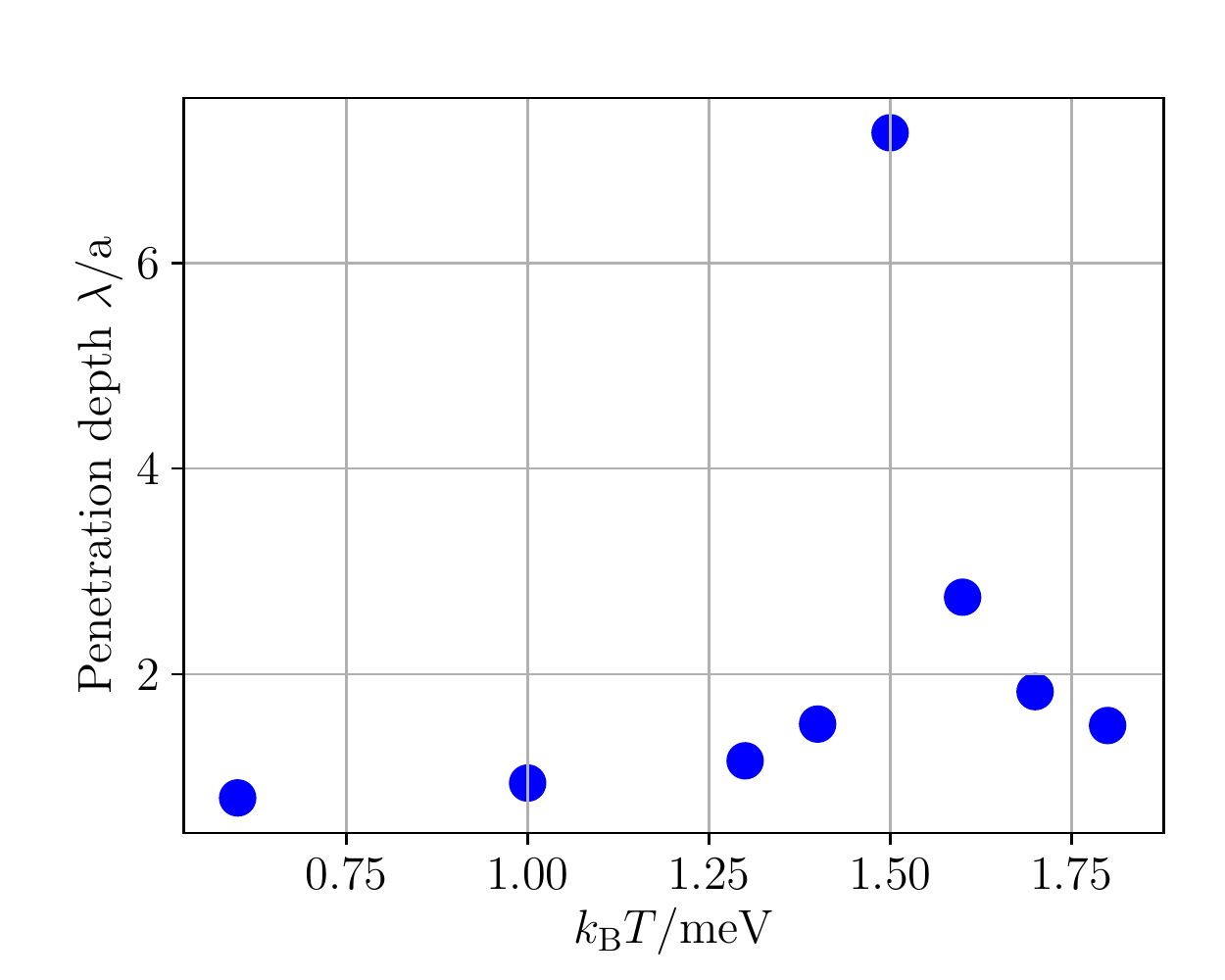} 
 \caption{%
    Temperature dependent penetration depth of the magnetic order, averaged from fitting exponential decays from the left and the right interface. Error bars are smaller than the symbol sizes.
  }
  \label{fig:PenetrationDepth}
\end{figure}

Analyzing the magnetization profiles further we find an enhanced difference from the bulk value closer to the interfaces.
The magnetization decays exponentially from the high value of the outer layer to the low value in the middle of the central layer. The according temperature-dependent decay constant which quantifies the penetration depth of the magnetic order is shown in \cref{fig:PenetrationDepth}.
These data, with a peak at the critical temperature, clearly demonstrate the occurrence of a critical behavior. From these observation we conclude that the magnetic order of the outer layers with the higher coupling constant penetrates into the central layer -- another signature of the magnetic proximity effect.
The length scale of this proximity effect is the correlation length of the system -- a quantity which in our case is only of the order of a few lattice constants though it should diverge at the critical temperature. Furthermore, its value might be much larger in materials with a larger range of the exchange interaction, beyond nearest neighbors. 

This proximity effect can also be observed in the monolayer-dependent magnetic susceptibility,
\begin{align}
  \chi^\text{FM}_{zz}(z) & = \frac{N}{\kBT}  \Big( \langle m_z(z) M_z \rangle - \langle m_z(z) \rangle \langle M_z \rangle \Big)  \label{eq:def_layerDepSuscep}
\end{align}
with the layer magnetization $m_z(z)$ and the total magnetization $M_z$ of the trilayer.
Note that this statistical definition equals the response-function definition $\chi^\text{FM}_{zz}(z) = \nicefrac{\del m_z(z)}{\del B_z}$, for a homogeneous magnetic field $\vec{B} = B_z\vec{e}_z$.

\begin{figure}
  \centering
  \includegraphics[width=0.9\linewidth]{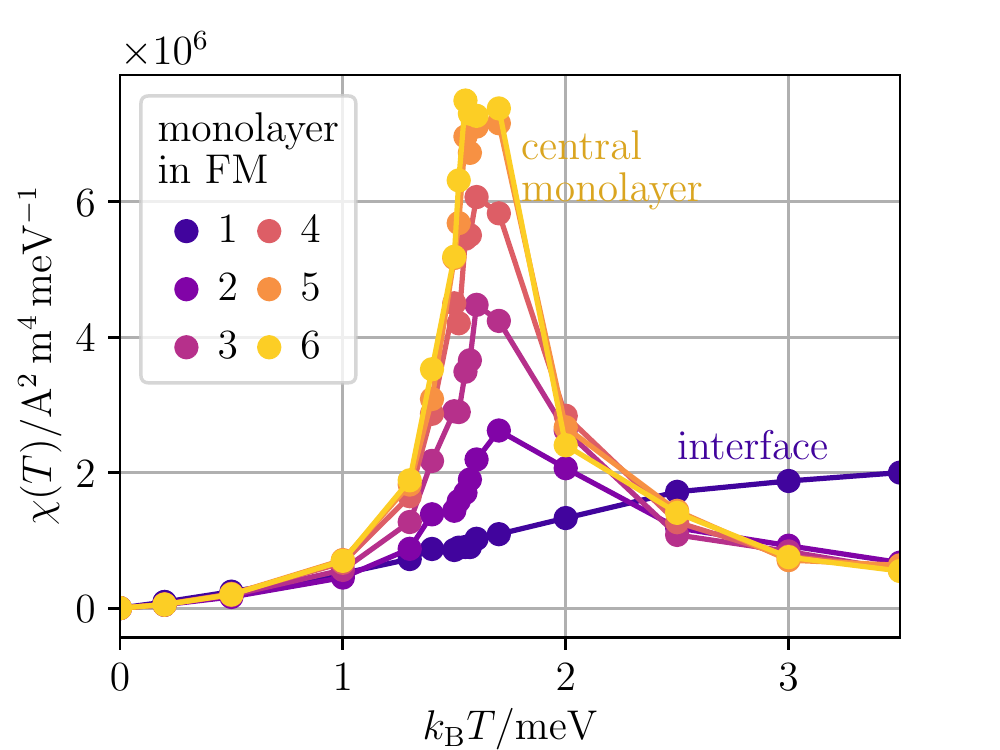}
  \caption{%
    Temperature dependence of the monolayer-wise susceptibility of the central layer of the FM-FM-FM exchange trilayer.
    For this calculation a thinner central layer of eleven atomic monolayers is used.
    The first layer (blue line) is directly at the interface, and the highest number (yellow line) denotes the monolayer in the middle of the central layer.
  }
  \label{fig:fmfmfmthinequilibriumlayeredsus}
\end{figure}

As shown in \cref{fig:fmfmfmthinequilibriumlayeredsus}, the critical behavior of the susceptibility in the central layer is suppressed, especially for those monolayers close to the interface (blue line).
In the middle of the central layer, there remains a maximum of the susceptibility around the expected critical temperature of the central layer as a reminiscent of the critical behavior of the corresponding bulk system.


\subsection{Comparison to a FM-lAFM-FM Trilayer}  \label{subsec:FM-lAFM-FM}
In the following, our previous results for a purely ferromagnetic trilayer are compared to the FM-lAFM-FM setup, with an antiferromagnet in the central layer. In the latter case, the order parameter of the central layer B is the \Neel{} vector $\mean{n_z} = \frac{1}{2} \left(\langle m_z^\uparrow \rangle - \langle m_z^\downarrow \rangle \right)$, where $m_z^{\uparrow\downarrow}$ are the corresponding sublattice magnetizations. In the outer layers A and C, the normal magnetization remains the order parameter as before.

\begin{figure}
  \centering
  \includegraphics[width=1.0\linewidth]{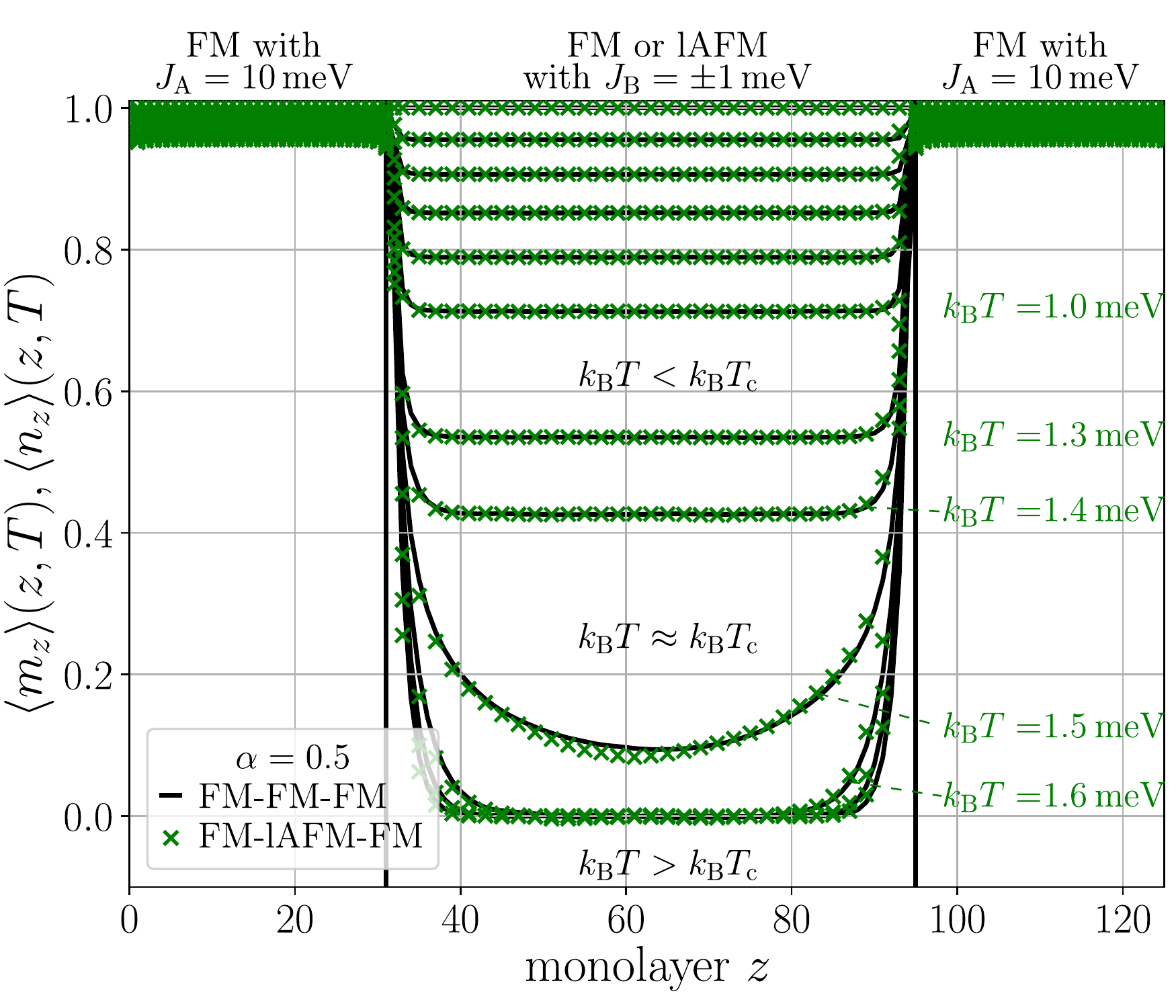}
  \caption{%
    Order parameter profiles for a FM-FM-FM trilayer (black solid line) compared to a FM-lAFM-FM trilayer (green crosses) with coupling ratio $\nicefrac{J_\mathrm{B}}{J_\mathrm{A}} = \num{0.1}$ for different temperatures.
  }
  \label{fig:FM-AFM}
\end{figure}

In \cref{fig:FM-AFM} the spatial- and temperature-dependent order parameter profiles of the two trilayers are compared (symbols for the AFM and lines for the FM).
Interestingly, they do not show any significant difference.
This is due to the fact that equilibrium properties result solely from the Hamiltonian of the system.
For the case of a two-sublattice AFM (with sublattices $\vec{S}^{\mathrm{A},\mathrm{B}}$) with only nearest-neighbor interaction there exists a transformation $J \mapsto -J$, $\vec{S}^\mathrm{B} \mapsto -\vec{S}^\mathrm{B}$ which maps the system to the corresponding ferromagnet.
The Hamiltonian is symmetric with respect to this transformation and, hence, the profiles are equal in equilibrium.
Note, however, that the argument above is solely classical and quantum corrections may lead to additional effects where equilibrium properties of FMs and AFMs deviate.

However, even in the central layer -- including its proximity effect --  we observe the same behavior for the \Neel{} vector of the central AFM in the FM-lAFM-FM trilayer as for the magnetization of the central FM in the FM-FM-FM trilayer: This is quite surprising since it means that a FM can generate even antiferromagnetic order via a proximity effect.
Looking at the exchange fields, however, it becomes clear that -- because of the fully uncompensated interface -- the nearest-neighbor exchange interaction of the FM acts on the layered AFM as a field that induces layer-wise the same order as in a FM.
Nevertheless, as we will show in the following, the magnon spectra in the two investigated trilayers will be affected differently by the proximity effect.


\subsection{Magnon Spectra}  \label{subsec:MagnonSpectra} 
In a further step, the magnonic spectra are calculated by a Fourier transform of the $N$ spins in time,
\begin{equation}
  \hat{S}_l(\omega) = \frac{1}{\sqrt{2\pi}} \int \left[ S_{l,x}(t) -i S_{l,y}(t) \right] e^{-i\omega t} \,\dd t,
\end{equation}
where the spin-wave amplitude for our easy-axis magnets is given by the $x$- and $y$ component of the spins.
For our numerical study, this property is calculated by a fast Fourier transform on discrete instances in time.
The frequency- and temperature dependent amplitude $\mathcal{I}(\omega,T)$ is then calculated as an average over all lattice sites 
\begin{equation}
  \mathcal{I}(\omega,T) = \frac{1}{N} \sum_l  \abs{ \hat{S}_l(\omega) }^2. 
\end{equation}

This quantity is proportional to the magnon number $n(\omega,T) = \mathrm{DOS}(\omega,T) \cdot f(\omega,T)$, where $\mathrm{DOS}$ is the density of states per volume and $f$ the magnon distribution function, in our classical spin model given by the Rayleigh-Jeans distribution $f(\omega,T) = \nicefrac{\kBT}{\hbar \omega}$.
Consequently, the quantity $\mathcal{I}(\omega,T)$ corresponds to a measurement of the magnon spectra, for instance by Brillouin light scattering \cite{Hillebrands00_ReviewBLSMultilayers}.

\begin{figure*}
  \centering 
  \includegraphics[width=0.48\linewidth]{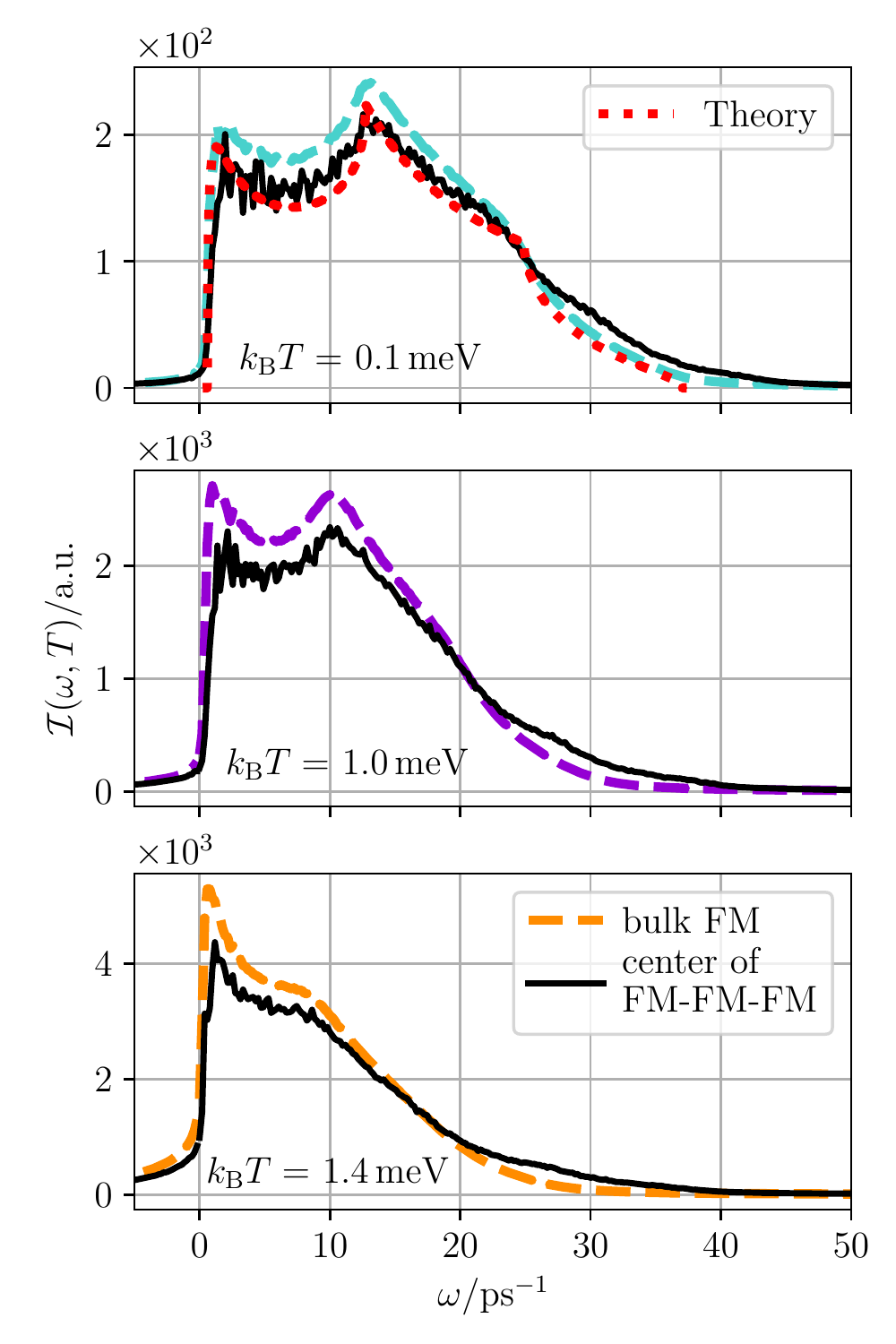}
  \includegraphics[width=0.48\linewidth]{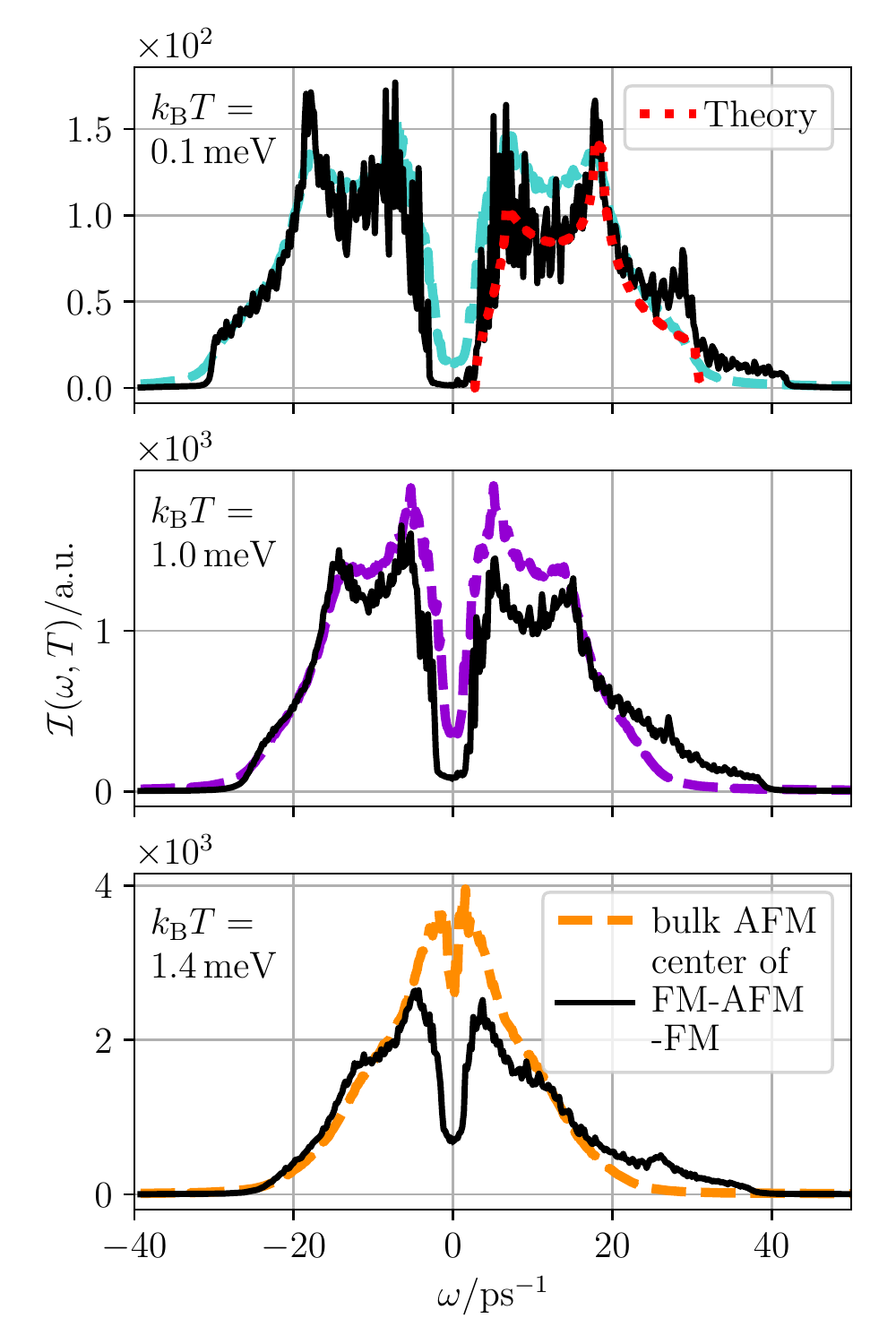}
  \caption{%
    Magnon spectra of the central layer (solid black lines) of a FM-FM-FM trilayer (left) and of a FM-lAFM-FM trilayer (right) with coupling ratio $\nicefrac{J_\mathrm{B}}{J_\mathrm{A}} = 0.1$ at different temperatures, compared to the according bulk spectra (colored dashed lines).
    For the lowest temperature (top graphs), a theoretical curve calculated in the limit of low temperatures is added.
  }
  \label{fig:MagnonSpectra}
\end{figure*}

\Cref{fig:MagnonSpectra}, left panel, shows the magnon spectra for the central layer of the FM-FM-FM trilayer (black solid lines) compared to a bulk ferromagnet (colored dashed lines) for increasing temperatures (top to bottom). Correspondingly, the right panel depicts the FM-lAFM-FM trilayer case in the same coloring.
Despite the fact that -- as shown before -- the two different order parameters follow exactly the same behavior, the spectra differ. The ferromagnet has only a single magnon branch with positive frequencies, whereas the antiferromagnet has two of opposite sign. Furthermore, the dispersion relation and, hence, the density of states are different \cite{Cramer18_SEEAcrossAFM}.

To reveal distinctive features we compare the trilayer spectra to the corresponding bulk spectra.
The figure depicts also the numerical bulk spectra (calculated by simulations of a pure bulk system) and for low temperatures the theoretical bulk spectra, calculated from the dispersion relations out of linear spin-wave theory.
These dispersion relations for a three dimensional simple cubic ferromagnet or layered antiferromagnet respectively read \cite{Cramer18_SEEAcrossAFM,eriksson_bergman_bergqvist_hellsvik_2017}
\begin{align} 
  & \frac{\mu_s}{\gamma} \omega_\text{FM}(\vec{k}) = 2d_z + 2J \sum_{ \mathclap{ \Theta\in \{x,y,z\} } }\left[1 - \cos(ak_\Theta)\right] 
\end{align}
and 
\begin{align} 
  & \frac{\mu_s}{\gamma} \omega_\text{lAFM}(\vec{k}) =  \nonumber \\
     & \pm \sqrt{ \Big[ 6|J| + 2d_z - 2|J| \smash{ \sum_{ \mathclap{ \Theta\in\{x,y\} } } } \cos(ak_\Theta) \Big]^2 - \Big[ 2|J|\cos(a k_z) \Big]^2}.
\end{align}
For the layered AFM, one has to include contributions from the antiferromagnetic coupling between the layers (along the $z$ direction) and ferromagnetic coupling within the layers ($x$-$y$ direction).
From these dispersion relations follows by integration the density of states and multiplied with the Rayleigh-Jeans distribution the theoretical curves in \Cref{fig:MagnonSpectra}.

Comparing trilayer systems and bulk we find distinct features at high and low frequencies.
High frequencies are around the maximal frequencies of the spectra of the central layer. These maximal frequencies can be read from the dispersion relations and they are $\omega_\text{max}^\text{FM} \approx \SI{36}{\per\ps}$ for the FM and $\omega_\text{max}^\text{lAFM} \approx \SI{31}{\per\ps}$ for the lAFM.
Remarkably, there are occupied states above this upper band edge.
These are most likely magnons from the outer FM layers, which have a ten times larger frequency range due to the higher coupling constant.
These magnons can penetrate the central layer in the form of evanescent waves. Consequently, 
even within the allowed frequency regime of the central layer, there are more high-frequency states occupied in the central layer than in the according bulk magnet. In the spectrum of the lAFM this manifests even as a peak at $\omega \approx \SI{28}{\per\pico\s}$.

In the low-frequency regime, there are significant deviations from a pure bulk spectrum. Not only is the amplitude here much lower in the central layer, also the position of the first maximum appears to be located at slightly higher frequencies. For the ferromagnetic trilayer we conclude that the low-frequency magnons can easily leave the central layer.
Since ferromagnetic magnons reduce the overall magnetization, the absence of magnons leads to the observation of an increased magnetization as compared to the bulk value.
This is perfectly in accordance with the observation from the order parameter curves \cref{subsec:FM-FM-FM}.

For the lAFM the resulting picture is more complicated, since only magnons with positive frequency can propagate into the outer FM layers, affecting the spectra asymmetrically even though magnons with negative frequencies can still migrate into the FM as evanescent waves. Indeed, we find a slight asymmetry with respect to positive and negative frequencies.
However, this asymmetry is not due to this effect alone, since, because of the odd number of monolayers within the lAFM, one of the sublattices is favoured and therefore also one of the branches.

A closer look also reveals further features: the lAFM has a temperature-dependent band gap.
The central layer of the trilayer exhibits the same effect, but close to the critical temperature, e.g.\ for $\kBT = \SI{1.4}{\milli\eV}$, the central layer still has a visible band gap, whereas in a bulk it has essentially vanished.
Thus it seems that the outer FM layers effectively cool the central lAFM, stabilizing the magnetic order and therefore weaken the effect of the vanishing bandgap at the critical temperature. 


\section{Conclusion}
We investigate and compare magnetic proximity effects in a FM-FM-FM and a FM-lAFM-FM exchange trilayer numerically using three different approaches. 
For spatially resolved and temperature-dependent order parameter profiles we show that magnetic order can be induced from the outer layers with higher critical temperature into the central layer with lower critical temperature. This is even true for a central antiferromagnetic layer and the order parameter profiles are the same for both types of trilayers.
In addition we studied the susceptibility profiles, finding a suppressed critical behavior in the vicinity of the interface as further signature of the magnetic proximity effect.

Most interestingly, magnon spectroscopy uncovers additional features, which could be summarized as  magnonic proximity effects: in the central layer there is a magnon occupation above the allowed frequency range and an additional peak close to the upper band edge of the AFM can be observed.
These effects are due to high-frequency spin waves from the outer layers with higher exchange coupling, which penetrate the central layer as evanescent modes. 
Nevertheless, the overall magnon number is lower -- a cooling effect due to the influence of the outer layers -- and the temperature dependence of the frequency gap is weakened.
Most importantly, the magnon spectrum of the central AFM becomes asymmetric since in the outer ferromagnetic layers only one polarization is allowed, an effect that was already exploited in a magnonic spin valve \cite{Cramer18_SpinValve}.
Our findings thus contribute to the understanding of magnetic equlibrium and spin transport phenomena in magnetic bi- and trilayers, especially at higher temperatures, approaching the critical temperature of one of the layers \cite{Cramer18_SEEAcrossAFM,Goennenwein_2018_MRatNeelTemp}.

%
%

\bibliography{Literature}

\end{document}